*Article*

# Cybersickness in Virtual Reality Questionnaire (CSQ-VR): A Validation and Comparison against SSQ and VRSQ.

**Panagiotis Kourtesis [1,2]\*, Josie Linnell[2], Rayaan Amir [2], Ferran Argelaguet[3] and Sarah E. MacPherson[2]**

[1] Department of Psychology, National and Kapodistrian University of Athens; pkourtesis@psych.uoa.gr
[2] Department of Psychology, University of Edinburgh; name.surname@ed.ac.uk
[3] Inria, Univ Rennes, CNRS, IRISA; ferran.argelaguet@inria.fr
\* Correspondence: pkourtesis@psych.uoa.gr

**Abstract:** Cybersickness is a drawback of virtual reality (VR), which also affects the cognitive and motor skills of the users. The Simulator Sickness Questionnaire (SSQ), and its variant, the Virtual Reality Sickness Questionnaire (VRSQ) are two tools that measure cybersickness. However, both tools suffer from important limitations, which raises concerns about their suitability. Two versions of the Cybersickness in VR Questionnaire (CSQ-VR), a paper-and-pencil and a 3D –VR version, were developed. Validation and comparison of CSQ-VR against SSQ and VRSQ were performed. Thirty-nine participants were exposed to 3 rides with linear and angular accelerations in VR. Assessments of cognitive and psychomotor skills were performed at baseline and after each ride. The validity of both versions of CSQ_VR was confirmed. Notably, CSQ-VR demonstrated substantially better internal consistency than both SSQ and VRSQ. Also, CSQ-VR scores had significantly better psychometric properties in detecting a temporary decline in performance due to cybersickness. Pupil size was a significant predictor of cybersickness intensity. In conclusion, the CSQ-VR is a valid assessment of cybersickness, with superior psychometric properties to SSQ and VRSQ. The CSQ-VR enables the assessment of cybersickness during VR exposure, and it benefits from examining pupil size, a biomarker of cybersickness.

## 1. Introduction

Virtual Reality (VR) is a promising technology that facilitates applications in many areas such as education [1], professional training [2], cognitive assessment [3], mental health therapy [4], and entertainment [5]. Nevertheless, beyond the advantages that VR brings to these fields, a limitation of VR is the presence of cybersickness that affects a percentage of users [6]. Cybersickness symptomatology includes nausea, disorientation, and oculomotor symptoms. Although there are similarities between cybersickness and simulator sickness, cybersickness differs from simulator sickness in terms of the frequency and severity of the types of symptoms [7]. Specifically, users experiencing cybersickness report increased general discomfort due to nausea and disorientation-related symptoms [7]. Cybersickness differs also from motion sickness, as cybersickness is triggered by visual stimulation, while motion sickness is triggered by actual movement [8].

Although there is not a comprehensive theoretical framework for cybersickness, the most frequent and predominant one is the sensory conflict theory [6], [8], [9]. This theoretical framework suggests that cybersickness symptomatology stems from a sensorial conflict between the vestibular (inner ear) and the visual system [6], [9]. In simple terms, the perception of postural balance relies on a combination of visual, vestibular, and proprioceptive input. Conflicting motion perception cues of the visual, proprioception, and vestibular systems are postulated to cause cybersickness. The technological reason for this conflict is vection, an illusory sense of motion, which occurs in VR. Vection is one of the main reasons for experiencing cybersickness in VR [10], [11]. Specifically, motions such as linear and angular accelerations appear to induce cybersickness in the user.

*1.1. Cybersickness, Cognition, and Motor Skills*

Beyond the obvious decrease in the user's experience in VR, cybersickness may also negatively affect the cognitive and/or motor performance of the user. Given that VR is used for applications that require intact cognitive and motor abilities (e.g., educational, research, clinical, and training applications), the presence of cybersickness has serious consequences for the implementation of VR in these applications. Recent systematic reviews of the literature suggest that cybersickness may substantially, yet temporarily, decrease the cognitive and/or motor performance of the user in immersive VR studies [12]–[14]. Dahlman et al. [15] postulated that motion sickness significantly decreases users' verbal working memory. Comparably, in immersive VR, Varmaghani et al. [16] conducted a study (N = 47), where the participants formed two groups: a VR group (N = 25) and a control group (N = 22; playing a board game). The results indicated that the VR group did not show an increase in visuospatial processing ability, while the control group did. Thus, this outcome postulated that cybersickness affects visuospatial processing and/or learning ability.

In another study, Mittelstaedt et al. [17] examined cybersickness and cognition (reaction time, spatial processing, visuospatial working memory, and visual attention processing) in pre- and post-sessions in VR. The findings showed that cybersickness modulated slower reaction speed, and prevented an expected improvement in visual processing speed [17]. These results suggest that cybersickness has a negative effect on attentional processing and reaction times, while spatial abilities and visuospatial memory remain intact. In the same vein, the studies of Nalivaiko et al. [18] (N = 26) and Nesbitt et al. [10] (N = 24) examined the effect of cybersickness on reaction times. In both studies, reaction speed was substantially slowed. Interestingly, slower reaction times were significantly correlated with an increase in the intensity of cybersickness [10], [18], indicating that cybersickness intensity may be associated with temporary cognitive and/or motor decline. However, no study has examined whether cybersickness intensity predicts cognitive or motor decline. Finally, while the above studies support the notion that cybersickness may decrease cognitive and/or motor skills, they all evaluated cybersickness after VR exposure, while no study assessed cybersickness during exposure.

*1.2. Cybersickness Questionnaires*

The Simulator Sickness Questionnaire (SSQ) is a 4-point Likert scale that was designed to assess simulator sickness in aviators [19]. The SSQ is the tool that has been used most frequently as a measure of cybersickness due to exposure to VR [13]. However, simulator sickness differs from cybersickness symptomatology, where in the latter disorientation and nausea symptoms are more frequent and intense [7]. Thus, despite its use in VR studies, the SSQ is not specific to the cybersickness symptoms that a user may experience in VR. Indeed, a recent study showed that SSQ does not have adequate psychometric properties for evaluating cybersickness in VR [20]. However, there is a variant of SSQ, the VR sickness questionnaire (VRSQ), that was recently developed [21] using items directly derived from the SSQ. In the development and validation study of the VRSQ, researchers attempted to isolate the items of the SSQ which are pertinent to cybersickness [21]. Nevertheless, this development and validation study suffered from serious limitations. Firstly, the sample size was small (i.e., 24 participants), and the stimuli diversity was limited.

Notably, the factor analyses accepted only items pertinent to oculomotor and disorientation symptoms, while they rejected all the items pertinent to nausea (i.e., 7 items) [21]. The latter is very problematic since it is well established that nausea is the second (after disorientation) most frequent type of symptoms of cybersickness [7], [22]–[24]. Furthermore, both the SSQ and VRSQ examine symptoms after VR exposure (and not during) and produce scores that cannot be easily interpreted. Finally, when developing the SSQ and VRSQ, the available guidelines for designing and developing a Likert scale tool were not considered.

There is scientific consensus over the design of Likert scale questionnaires. The literature suggests that a 7-point Likert scale is substantially better than a 5-point (or less) one [25]–[28]. The 7-point design offers a greater variety of responses, which better captures the diversity of the individuals' views or experiences. Furthermore, combining numbers (e.g., 6) with corresponding text (e.g., strongly disagree) facilitates a better understanding of the differentiation amongst the available responses [25]–[28]. These suggestions have been considered and adopted in the development of the VR Neuroscience Questionnaire (VRNQ) and the CyberSickness in VR Questionnaire (CSQ-VR) [29]. The CSQ-VR is derived from the VR Induced Symptoms and Effects (VRISE) section of the VR Neuroscience Questionnaire (VRNQ), which has been found to have very good structural and construct validity [29]. Also, the VRISE section of the VRNQ has been validated against the SSQ and the Fast Motion Sickness Scale [30]. The advantages of the VRISE over the SSQ pertain to its short administration (only 5 items/questions) and the production of easily comprehensible outcomes [30]. However, the scoring of the VRISE is inverse (i.e., higher scores indicate milder symptom intensity). In addition, oculomotor symptoms were assessed by only one question in the VRISE.

The CSQ-VR was designed in line with the aforementioned guidelines (i.e., using a 7-point scale and combining text with numbers), while it also addressed the previous shortcomings (i.e., inverse scoring and one oculomotor question) of the VRISE section of the VRNQ. The CSQ-VR assesses the whole range of cybersickness symptoms, including nausea, disorientation, and oculomotor. There are two questions for each type of symptom. Each question is presented on a 7-item Likert Scale and the responses are offered as combined text and numbers, ranging from "1 - absent feeling" to "7 - extreme feeling". The CSQ-VR produces a Total Score and three sub-scores: Nausea, Disorientation, and Oculomotor. Each sub-score corresponds to a type of symptoms and is calculated by adding the two corresponding responses. The Total Score is the sum of the three sub-scores. The design of the CSQ-VR allowed the advantages associated with the VRISE of the VRNQ (i.e., very short administration, easy and interpretable scoring, comprehensible questions and responses, and examination of all types of cybersickness' symptoms) to be maintained and the weaker aspects (i.e., the addition of one more oculomotor question and positive scoring, where larger numbers indicate stronger symptoms) to be improved. Finally, the CSQ-VR was not only developed in a paper-and-pencil form, but also in a 3D form that can be used in any virtual environment to examine cybersickness while the user is in VR. This VR version of the CSQ-VR also benefits from eye-tracking for measuring gaze fixations and pupil size (i.e., pupillometry). Since pupil size has been associated with negative emotions [31], pupillometry may offer a physiological metric of cybersickness intensity.

*1.3. Research Aims*

This study aims to examine the validity of the paper-and-pencil version and the VR version of the CSQ-VR in detecting and evaluating cybersickness symptoms. The validation will be examined against the SSQ and the VRSQ, which are considered valid tools of cybersickness. Furthermore, as there is an association between cybersickness and cognitive and motor performance, this study offers a comparison among CSQ-VR (both versions), SSQ, and VRSQ in detecting temporary cognitive and/or motor decline due to cybersickness. The VR version of CSQ-VR is expected to facilitate an ongoing examination of cybersickness while the user is immersed. Finally, the utility of pupillometry in predicting cybersickness intensity will also be explored.

## 2. Materials and Methods

*2.1. Virtual Environment Development*

The virtual environment was developed using the Unity3D game engine. The interactions with the environment were developed using the SteamVR SDK. Since gaming experience may modulate task performance [3], the virtual hands/gloves of SteamVR SDK were used to ensure an ergonomic and effortless interaction. Notably, none of the interactions required button presses. Instead, interactions were facilitated by simply touching the object (initial selection) and continuously touching the object (to confirm the selection). In addition, SteamVR virtual hands/gloves do not represent any gender or race, so their utilization prevents confounding effects from these variables [32].

To ensure understanding and seamless completion of the tasks, users received instructions in video, audio, and written form. For each task's instructions, audio clips with neutral naturalistic voices were produced using the Amazon Polly. The audio feedback was spatialized using the SteamAudio plugin. The SRapinal SDK was used for eye-tracking and facilitating pupillometry. Finally, randomization of the experimental blocks within and between participants, and extraction of the data into a CSV file, as well as the facilitation of the experimental design and control was achieved using bmlTUX SDK [33].

2.1.1. Linear and Angular Accelerations in VR

Based on the relevant literature, linear and angular accelerations are efficient in inducing significant cybersickness symptoms in users in a relatively short time (e.g., 5-10 mins) [10], [12], [13], [18], [24], [29], [34]. Correspondingly, a ride of 5 mins was developed. Since the ride had to be repeated three times (i.e., a total 15 min ride) for each participant, a 5 min duration was preferred. The ride was designed as an animation of the platform that the user was standing on (see Figure 1). The direction of the motion was always forward (except in the last stage; see reversed z-axis). The movements of the platform were similar to those of a roller coaster. The ride included the following accelerations in this specific order: 1) linear (z-axis), 2) angular (z and y axes), 3) angular (z, x, and y axes), 4) angular (roll axis), 5) extreme linear (z-axis), 6) angular (yaw axis), and 7) extreme linear (y-axis followed by reversed z-axis). The environment had simple black-and-white surroundings (see Figure 1). This background was used to ensure that the symptoms are strictly induced by vection, and not due to other reasons such as intense colours. Also, having the squared/tiled design offered cues to the participants for perceiving vection and altitude changes.

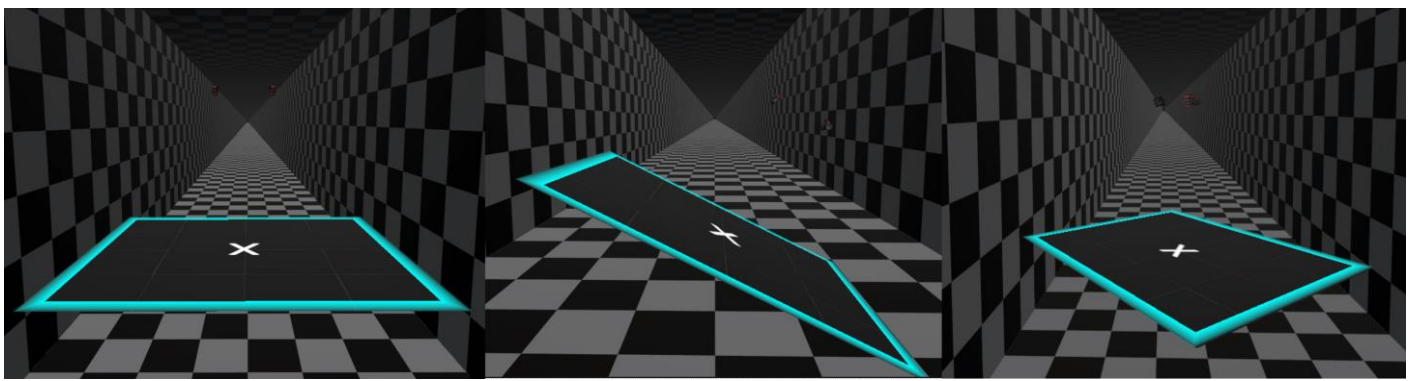

**Figure 1.** Examples of Linear (Left) and Angular (Centre & Right) Accelerations during the Ride.

*2.2. Cognitive and Psychomotor Skills' Assessment*

The aims of this study required the examination of cybersickness, cognition, and motor skills to be repeated, while the user is immersed in VR. For these reasons, immersive VR versions of well-established tests were developed. For the development of these VR cognitive and psychomotor tasks, the specific design and development guidelines and recommendations for cognitive assessments in immersive VR were followed [35].

2.2.1. Verbal Working Memory

A VR version of the Backward Digit Span Task (BDST; [36]) was developed and used. The VR BDST requires participants to listen to a series of digits and remember and recall them in the reverse order of their presentation. For example, when the digits were 2, 4, and 3, then participants had to respond in the reverse order (i.e., 3, 4, 2). Therefore, the first step involved listening to the digits. After this step, a keypad appeared in front of the participants. Using the keypad, users had to provide the digits in reverse order. To indicate a number, participants had to touch the white box button displaying the equivalent number (see Figure 2). Continuous touch of the button for one second confirmed the response. After confirmation, if the response was correct, the button turned green and made a positive sound. In contrast, if the response was incorrect, then the button turned orange and made a negative sound. When a mistake was made or all the digits were provided correctly, the trial ended. In every second successful trial, the length of the digit sequence was increased. When the participant made two subsequent mistakes within the same digit sequence length (e.g., 3 digits), or when they finished the last trial (i.e., second trial with a sequence length of 7 digits), then the task ended. The Total Score of the VR BDST was determined by adding together the total number of correct trials and the highest digit sequence length that at least one trial was successfully performed. A video displaying the task and its procedures can be found here: [Link to the Video](#).

2.2.2. Visuospatial Working Memory

Visuospatial working memory was assessed using the Backward Corsi Block Test (BCBT) [37]. A VR version of the BCBT was developed. This task consists of 27 white boxes, where each one is placed in a different position, based on the x, y, and z axes. Nevertheless, only nine boxes out of the 27 possible boxes were shown to the participants at one time (see Figure 2). The 9 boxes were presented at the beginning of each trial. Then, a number of these boxes (depending on the current sequence length) were randomly presented (turning blue and making a bell sound) in sequential order, with each box presented for one second. After the presentation of the sequence, participants had to select the boxes in reverse order. Participants had to touch a cube (the cube turned blue on touch) and keep touching it for one second to select the cube. When a cube was selected, it either turned green and made a positive sound (i.e., correct response), or it turned orange and made a negative sound (i.e., an error). The trial ended when participants either made a mistake, or they correctly selected all the targets in their reverse order. The sequence lengths were initially two boxes, with two trials for each length. The number of boxes in the sequence was increased by one box when at least one of the two trials of the same length/span was correct. When the participant incorrectly recalled two sequences of the same length, the task was ended. Equally, when the second trial of the last length/span (i.e., 7 cubes) was performed, the task ended. Sequence lengths increased up to seven cubes. The Total Score is the sum of the span (the longest correct sequence length) and the total number of correct sequences. A video displaying the task and its procedures can be found here: [Link to the Video](#).

2.2.3. Psychomotor Skills

To assess reaction times, a VR version of the Deary–Liewald Reaction Time (DLRT) task [38] was developed and used. The DLRT encompass two tasks. One task assesses simple reaction time (SRT), and the other task examines choice reaction time (CRT). For the SRT task, participants have to observe a white box and touch it as soon as the box changes colour to blue (see Figure 2). There are 20 trials/repetitions in the SRT task. In the CRT task, there are four boxes, which are aligned horizontally (see Figure 2). Randomly, one of the four boxes changes its colour to blue. When the box turns blue, participants are required to touch the box as fast as possible (see Figure 2). The CRT task includes 40 trials/repetitions. For both SRT and CRT, the participants are instructed to touch the boxes as fast as possible using the most convenient hand. There was a practice session at the start of both the SRT and the CRT to ensure that the instructions were understood by participants. A video displaying the task and its procedures can be found here: [Link to the Video](#).

As in the original version, the SRT produces a score that is the average reaction time across the 20 trials. Similarly, the CRT produces a score that is the average reaction time across the 40 trials, for the correct responses only. However, in addition, given that the VR version of the CRT is enhanced by eye-tracking, the time required to attend to the target was also measured (Attentional Time; i.e., the time from the appearance of the target until the gaze of the user falls on it). Also, eye-tracking facilitated the calculation of the time required to touch the target once it had been attended to (Motor Time). Finally, similarly to the original version, the overall time between the target's presentation and its selection (Reaction Time) was also calculated. Thus, the VR version of CRT produces three scores:

1) the Reaction Time (RT) to indicate the overall psychomotor speed
2) the Attentional Time (AT) to indicate the attentional processing speed
3) the Motor Time (MT) to indicate the movement speed

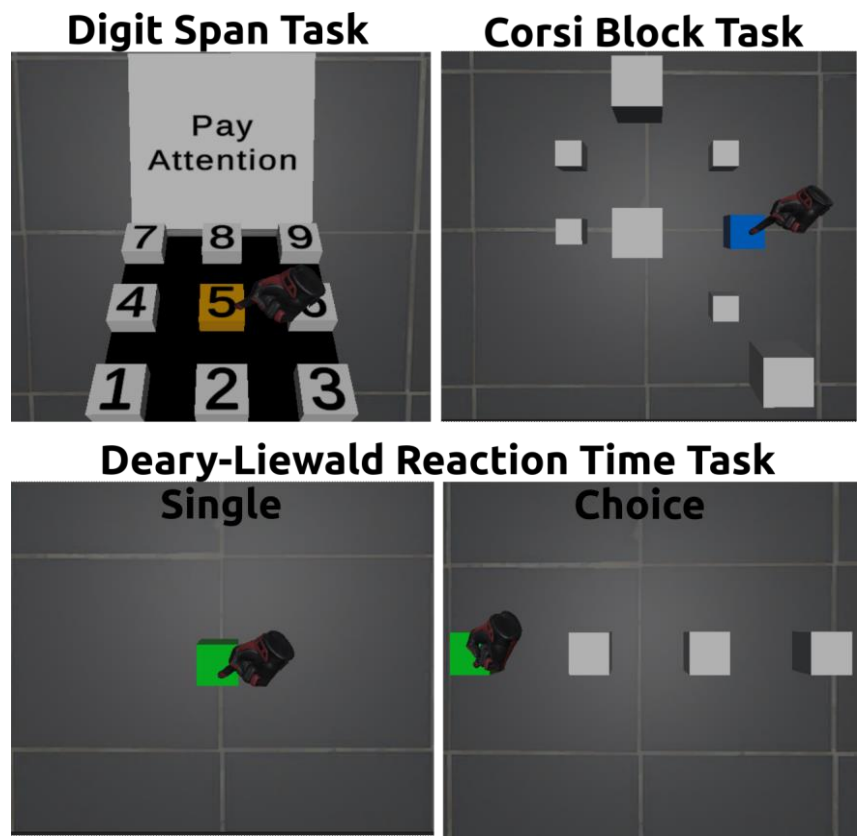

**Figure 2.** Digit Span Task (Upper Left), Corsi Block Task (Upper Right), and Deary-Liewald Reaction Time Tasks (Bottom).

*2.3. Cybersickness Questionnaires*

The Motion Sickness Susceptibility Questionnaire (MSSQ) [39] was completed prior to enrolment, to reduce the likelihood of a participant experiencing severe symptoms of cybersickness. The MSSQ is a 3-point Likert scale with 18 items/questions examining the experience of motion sickness using diverse means of transport. Nine items refer to the experience of motion sickness as a child and the other nine items as an adult. The nine questions hence are repeated in both sections. The MSSQ produces three scores: a Child score; an Adult score; and a Total score, which is the addition of the previous two scores.

The SSQ and the VRSQ were administered pre- and post-exposure to VR to assess the intensity of cybersickness symptoms. Both SSQ and VRSQ are 4-point Likert scales. The SSQ was developed for individuals that are trained using simulators (e.g., aviators) [19]. The SSQ has 16 questions, which are grouped under three categories: Nausea; Disorientation; and Oculomotor. Four scores are produced, one for each category, and a Total score. The calculation of the scores is made by a formula offered by the developers of the SSQ [19]. The maximum score is 100 for each category, and 300 for the Total Score. On the other hand, the VRSQ derives from the SSQ, and it contains 9 items (i.e., approximately half of the SSQ items), which are grouped under two categories: Disorientation and Oculomotor (i.e., the Nausea items were excluded) [21]. The VRSQ produces three scores, one for each category, and a Total Score, which is the sum of the two sub-scores divided by two. The maximum score for each sub-score is 100. As discussed above (see subsection 1.2), while both the SSQ and VRSQ appear to be valid tools, they suffer from certain limitations:

- The SSQ is not specific to cybersickness, while the frequency and intensity of symptoms substantially differ in simulator sickness and cybersickness.
- The VRSQ does not consider nausea symptoms, while nausea symptoms are the second most frequent type of symptoms in cybersickness.
- The VRSQ validation was performed in a study with a small sample size and a limited diversity of stimuli.
- Both the SSQ and VRSQ, being 4-point Likert scales, were not designed in line with the design guidelines for Likert scale questionnaires.

2.3.1. Cybersickness in VR Questionnaire

The CSQ-VR is an improved version of the VRISE section of VRNQ. The VRISE section has been found to have very good structural validity in a study where participants were exposed to three diverse VR software and environments [29]. Also, the VRISE section of the VRNQ has been previously validated and compared against the SSQ [30]. The VRISE of VRNQ appeared superior to the SSQ due to its shorter administration time (i.e., 5 items instead of 16 items) and the enhanced interpretability of the scores (i.e., scores calculated by a simple addition, instead of a complex formula). However, the VRISE section of the VRNQ had only one item for Oculomotor symptoms, and the score was inversed (i.e., a higher score indicated a weaker intensity of that symptom). To address these limitations, the CSQ-VR was developed based on the VRISE section of the VRNQ. Comparably to the VRNQ, the CSQ-VR was designed and developed by following the design guidelines for Likert scales i.e., a 7-point Likert scale and combining text with numbers in the responses (see [25]–[28]). The CSQ-VR is a 7-point Likert scale that includes six questions for the assessment of the three types of symptoms of cybersickness, which form the respective sub-scores: Nausea, Vestibular, and Oculomotor. Each category includes two questions. The Total Score is the sum of the three scores, which a maximum score of 42 (14 for each sub-score). The paper-and-pencil version of the CSQ-VR can be found in Supplementary Material.

Moreover, a 3D version of the CSQ-VR has also been developed to assess cybersickness while the user is immersed in VR. A User Interface (UI) for the VR version of CSQ-VR was designed and developed. In the UI, the question appears in the upper area and the response (in red letters) is in the middle area. The users change their response, by touching the corresponding number, or sliding along the slider (see Figure 3). Furthermore, based on the established link between pupil size and affective/emotional state [31], eye-tracking was integrated to facilitate ophthalmometry and pupillometry. For measuring Fixation Duration, invisible eye-tracking targets were placed in front of the text, while their height and width were always matched to the displayed text per line (see Figure 3).

Moreover, the measurement of pupil size was continuous while the user responded to the CSQ-VR questions. Pupillometry enabled measurement of the average Pupil Size (right and left), which can be used as a physiological metric of negative emotion. Finally, a video showing the procedures of a questionnaire in VR can be found here: Link to the Video.

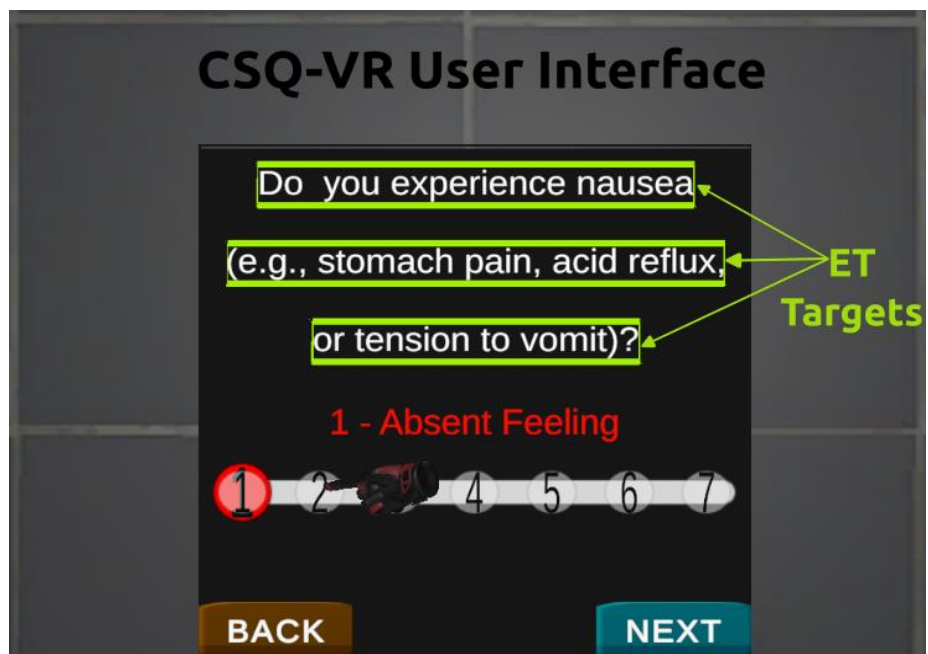

**Figure 3.** User Interface and Eye-Tracking (ET) Targets of the VR version of CSQ-VR. Note: Eye-tracking targets were not visible to the user.

*2.4. Participants and Procedures*

Thirty-nine participants (22 females, 17 males) were recruited with a mean age of 25.28 years [*SD = 3.25, Range = 22-36*] and mean education of 17.23 years [*SD = 1.60, Range = 13-20*]. The recruitment was performed via opportunity sampling, using the University of Edinburgh's internal mailing lists, alongside advertisements on social media. The study was approved by the School of Philosophy, Psychology, and Language Sciences (PPLS) Ethics Committee of the University of Edinburgh. Informed and written consent was obtained from all participants prior to their participation. Participants were compensated with £20 each for their time and effort.

The MSSQ was completed before enrolment to reduce the likelihood of severe symptoms following VR exposure. In line with the MSSQ author's suggestions [39], the 75th percentile was used as a parsimonious cut-off score for inclusion in the study. This allowed us to exclude the individuals who are susceptible to experiencing strong cybersickness symptomatology (i.e., the upper 25th percentile of the population). The included participants were then invited to attend the experiment. Upon arrival, participants were informed of the study's aims and procedures, and the adverse effects that they may experience. The participants then provided their informed consent in written form.

Firstly, an induction on how to wear the headset and use and hold the controllers was offered to every participant. An HTC Vive Pro Eye was used, which embeds an eye-tracker with a 120Hz refresh rate and a tracking accuracy of 0.5°-1.1°. Secondly, the participants provided their demographic data: age, sex, gender, education, dominant eye, VR experience, computing experience, and gaming experience, by responding to a questionnaire. The dominant eye was determined using the Miles Test [40]. Note that VR/computing/gaming experiences were calculated by adding the scores from two questions (6-item Likert Scale) for each one. The first question was pertinent to the participant's ability (e.g., 5 - highly skilled) to operate a VR/computer/game, and the second one was pertinent to the frequency of operating them (e.g., 4 - once a week).

Before VR exposure, participants responded to the CSQ-VR (paper version), SSQ, and VRSQ. Participants then were immersed in VR. Note that for the assessments and rides, participants were always in a standing position in the middle of the VR area (see the X mark in Figure 1). The first part included the tutorials, during which a video tutorial for each task was offered, alongside the corresponding verbal and written instructions. After each tutorial, the participant performed the corresponding task. This part formed the baseline assessment of each participant. The baseline assessment included: the VR version of CSQ-VR (Cybersickness), the verbal working memory task (BDST), the visuospatial working memory task (BCBT), and the reaction time task (DLRT; see Figure 2). After the baseline assessment, the first ride started. After each ride, the participants performed an assessment identical to the baseline (i.e., CSQ-VR, BDST, BCBT, DLRT). On top of the baseline assessment, the participants were exposed to three rides and three respective assessments. The whole procedure in VR lasted approximately 100 mins for each participant. After the VR session, participants responded to the CSQ-VR (paper version), SSQ, and VRSQ. Then, refreshments rich in electrolytes were offered to the participants. Moreover, the participants rested for 10-15 mins before leaving the premises. The participants were instructed to avoid driving and using heavy machinery for the rest of the day.

*2.5. Statistical Analyses*

Descriptive statistical analysis was performed to provide an overview of the sample. Reliability analyses were conducted to examine the internal consistency of the CSQ-VR. The recommended thresholds for Cronbach's $\alpha$ were used for interpreting the internal consistency (i.e., adequate = 0.6-0.7, good = 0.7-0.8, and very good = 0.8-0.95) [41]. Pearson's correlational analyses were performed to examine the validity of the CSQ-VR versions against the SSQ and the VRS post-exposure (i.e., after the VR session). Since the SSQ is considered the gold standard, and it also has a structure (i.e., three sub-scores; Nausea, Oculomotor, Disorientation) similar to CSQ-VR, the convergent validity (i.e., correlations) of CSQ-VR was assessed against SSQ. Receiver Operating Characteristic (ROC) and Area Under the Curve (AUC) analyses were performed to appraise the psychometric properties of CSQ-VR, SSQ, and VRSQ in detecting temporary cognitive and motor declines due to cybersickness. The thresholds of AUC > 0.7 and Metric Score > 1.5 were used in line with the respective recommendations for determining the suitability of the tool [42], [43]. The temporary decline was based on performance on the assessment after each ride. In agreement with the consensus of the American Academy of Clinical Neuropsychology for determining a substantial decrease in performance, two standard deviations from the mean were used [44]. Thus, when performance (i.e., score) on the assessment after the respective ride was 2 standard deviations from the mean of the baseline assessment, performance was defined as abnormal (i.e., temporary decline). Note that the two standard deviations had to indicate worse performance, thus, be greater for the reaction and motor times (i.e., slower reaction or motor speed) and smaller for the verbal and visuospatial working memory (i.e., poorer performance). Finally, the predictive ability of pupil size was examined by performing a mixed model regression analysis. The analysis was performed using the *Jamovi* statistical software (Descriptive Statistics, Reliability, ROC, and AUC analyses) [45], as well as *R* (transforming the data, plots design, correlation and regressions analyses) [46]. As the variables violated the normality assumption, we used the *bestNormalize* R package [47] to transform and centralize the data. The distribution of the data was then normal. The transformed data were used for the parametric analyses (i.e., correlations and mixed regression analysis). Furthermore, the psych (correlational analyses) [48], the *ggplot2* (plots) [49], and the *lme4* (regression analyses) [50] R packages were used for performing the respective analyses.

## 3. Results

The descriptive statistics of the sample are displayed in Table 1. Concerning the intensity of cybersickness symptoms, it can be observed that participants predominantly experienced moderate symptoms. There were no dropouts during the experiment. The descriptive statistics for the VR version of CSQ-VR, per experimental stage, are presented in Table 2.

**Table 1.** Descriptive Statistics

|  | Mean (SD) | Range | Max. Score |
|---|---|---|---|
| Sex (22F/17M) | - | - | - |
| Age | 25.28 (3.22) | 22 – 36 | - |
| Years of Education | 15.14 (5.18) | 13 – 20 | - |
| VR Experience | 2.67 (0.92) | 2 – 6 | 14 |
| Computing Experience | 10.36 (0.80) | 9 – 12 | 14 |
| Gaming Experience | 5.54 (2.97) | 2 – 12 | 14 |
| MSSQ Child Score | 4.69 (3.34) | 0 – 13.50 | 27 |
| MSSQ Adult Score | 3.91 (3.20) | 0 – 11.25 | 27 |
| MSSQ Total Score | 8.60 (5.23) | 0 – 20.13 | 54 |
| Pupil Size (mm) | 5.37 (0.90) | 3.70 – 8.32 | - |
| CSQ-VR (VR) Total Score* | 10.63 (4.97) | 6 – 28 | 42 |
| CSQ-VR (VR) Nausea Score* | 3.18 (1.56) | 2 – 9 | 14 |
| CSQ-VR (VR) Vestibular Score* | 3.66 (2.43) | 2 – 13 | 14 |
| CSQ-VR (VR) Oculomotor Score* | 3.79 (1.70) | 2 – 9 | 14 |
| CSQ-VR Total Score | 12.23 (4.96) | 6 – 27 | 42 |
| CSQ-VR Nausea Score | 3.51 (1.68) | 2 – 9 | 14 |
| CSQ-VR Vestibular Score | 3.97 (2.41) | 2 – 10 | 14 |
| CSQ- VR Oculomotor Score | 4.74 (1.81) | 2 – 10 | 14 |
| SSQ-Total Score | 67.24 (48.09) | 0 – 223.66 | 300 |
| SSQ-Nausea Score | 24.22 (22.09) | 0 – 95.40 | 100 |
| SSQ-Disorientation Score | 9.40 (9.98) | 0 – 44.88 | 100 |
| SSQ-Oculomotor Score | 33.62 (21.84) | 0 – 83.38 | 100 |
| VRSQ-Total Score | 19.17 (13.27) | 0 – 59.17 | 100 |
| VRSQ-Disorientation Score | 11.62 (13.27) | 0 – 60.00 | 100 |
| VRSQ-Oculomotor Score | 26.71 (15.63) | 0 – 58.33 | 100 |

*(VR) = VR version; Pupil Size measured while responding to the VR version of CSQ-VR.*

**Table 2.** Descriptive Statistics of the VR version of CSQ-VR per Experimental Stage

| Experimental Stage | CSQ-VR Scores* | Mean (SD) | Range | Max. Score |
|---|---|---|---|---|
| Baseline | Total Score | 7.59 (2.09) | 6 – 16 | 42 |
| | Nausea Score | 2.23 (0.54) | 2 – 4 | 14 |
| | Vestibular Score | 2.38 (0.85) | 2 – 6 | 14 |
| | Oculomotor Score | 2.79 (1.11) | 2 – 6 | 14 |
| Ride 1 | Total Score | 10.79 (4.35) | 6 – 24 | 42 |
| | Nausea Score | 3.41 (1.37) | 2 – 8 | 14 |
| | Vestibular Score | 3.97 (2.47) | 2 – 12 | 14 |
| | Oculomotor Score | 3.41 (1.41) | 2 – 8 | 14 |
| Ride 2 | Total Score | 11.87 (5.03) | 6 – 23 | 42 |
| | Nausea Score | 3.54 (1.57) | 2 – 8 | 14 |
| | Vestibular Score | 4.13 (2.56) | 2 – 12 | 14 |
| | Oculomotor Score | 4.21 (1.73) | 2 – 9 | 14 |
| Ride 3 | Total Score | 12.26 (6.19) | 6 – 28 | 42 |
| | Nausea Score | 3.54 (2.02) | 2 – 9 | 14 |
| | Vestibular Score | 4.15 (2.91) | 2 – 13 | 14 |
| | Oculomotor Score | 4.56 (2.00) | 2 – 9 | 14 |

*Scores of the VR version of CSQ-VR during the exposure to VR*

### 3.1. Reliability and Validity

Interpretation of the outcomes was based on the recommendations offered by Ursachi et al. [41]. Based on them, Cronbach's $\alpha$ of 0.6-0.7 is an acceptable score, 0.7-0.8 is a good score, and 0.8-0.95 is a very good score. The overall internal consistency of the questionnaire (i.e., the Total score's reliability) was evaluated by considering each questionnaire's sub-scores. The internal consistency of the sub-categories (i.e., the reliability of the sub-score) was examined by considering the respective items/questions. The reliability analysis revealed that all sub-scores of the CSQ-VR have good internal consistency (see Table 3). Specifically, the Total Score and the Vestibular sub-score showed very good internal consistency, while the Nausea and Oculomotor sub-scores revealed good internal consistency. However, the Cronbach's $\alpha$ of the Oculomotor sub-score was at the margins between good and adequate internal consistency.

Both the SSQ and VRSQ Total scores showed good internal consistency, however, both were substantially lower than the internal consistency of the CSQ-VR Total Score (see Table 3). The Nausea score of the SSQ showed an acceptable internal consistency, which is significantly lower than the almost very good internal consistency of the Nausea score of the CSQ-VR. The Disorientation score of the SSQ revealed marginally very good internal consistency, while the Disorientation score of the VRSQ indicated marginally good internal consistency. Both Disorientation scores (SSQ and VRSQ) were significantly lower than the almost excellent internal consistency of the CSQ-VR. Finally, the Oculomotor score of the SSQ showed good internal consistency that was higher than the marginally good internal consistency of the Oculomotor score of the CSQ-VR. On the other hand, the Oculomotor score of the VRSQ revealed adequate internal consistency. The Oculomotor score of the VRSQ and the Nausea score of the SSQ were the two scores that were below the parsimonious threshold of 0.7. Overall, the CSQ-VR appeared to have superior internal consistency compared to the SSQ and the VRSQ.

Table 3. Reliability (Internal Consistency) of CSQ-VR, SSQ, and VRSQ.

| Questionnaire | Scores | Cronbach's $\alpha$ |
|---|---|---|
| **CSQ-VR** | Total Score | 0.865 |
| | Nausea | 0.792 |
| | Vestibular | 0.934 |
| | Oculomotor | 0.704 |
| **SSQ** | Total Score | 0.810 |
| | Nausea | 0.676 |
| | Disorientation | 0.809 |
| | Oculomotor | 0.744 |
| **VRSQ** | Total Score | 0.806 |
| | Disorientation | 0.718 |
| | Oculomotor | 0.654 |

*The internal consistency was based on the sub-scores of the Total score, and the sub-scores were based on their respective items. Based on [41], Cronbach's α of 0.6-0.7 is acceptable, 0.7-0.8 is good, and 0.8-0.95 is very good.*

The scores of the CSQ-VR (of both versions) were significantly correlated with the corresponding scores of the SSQ. Overall, the analyses revealed moderate to strong correlations between the scores. The paper-and-pencil version of CSQ-VR was strongly associated with the SSQ (see Figure 4); especially, their total scores, as well as their oculomotor scores, revealed a very strong correlation between them. Although the correlations for the Nausea and Vestibular scores were weaker than those observed above, they were still strong correlations (see Figure 4). Similarly, the VR version of the CSQ-VR was strongly associated with the SSQ (see Figure 5). In particular, their total scores indicated a strong correlation between them. While the correlations for their sub-scores were weaker than those between the total score, they were still moderate-to-strong correlations (see Figure 5). These results postulate the convergent validity of both versions of the CSQ-VR. Also, given that all sub-scores were substantially associated, the construct validity of the CSQ-VR is strongly supported.

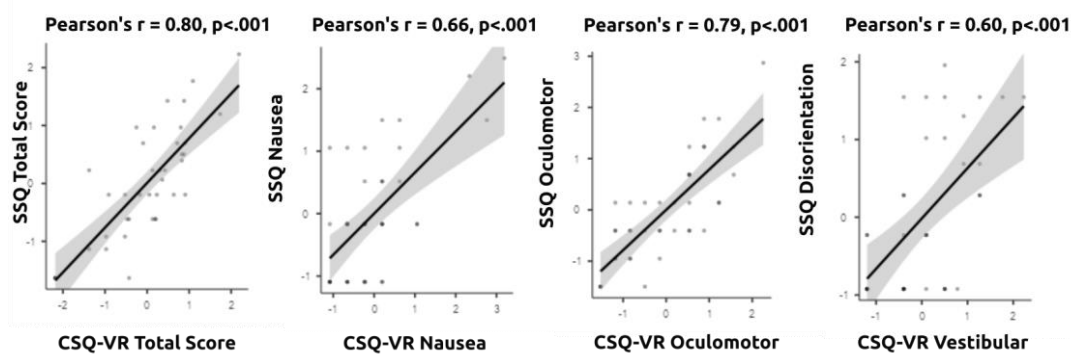

**Figure 4.** Correlations between the scores of the CSQ-VR (paper-and-pencil version) and the SSQ.

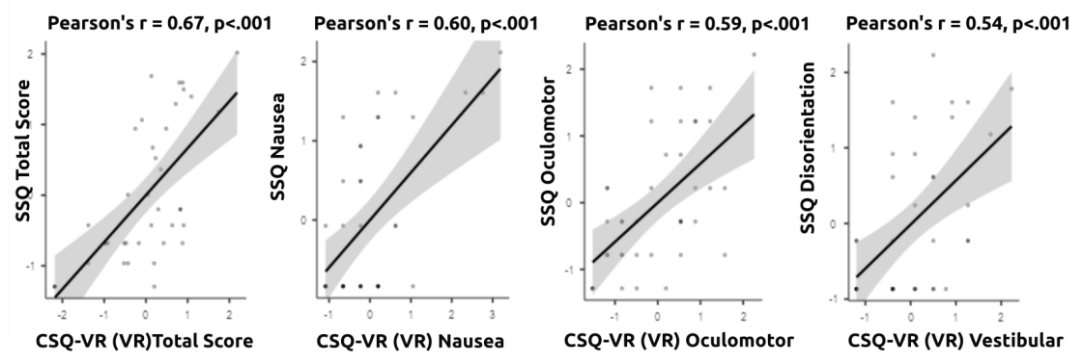

**Figure 5.** Correlations between the scores of the CSQ-VR (VR version) and the SSQ.

Furthermore, the scores for both versions of the CSQ-VR were strongly associated with the VRSQ scores (see Table 4). The total scores of the CSQ-VR versions showed the strongest correlations with the total score of the VRSQ. The oculomotor scores of the CSQ-VR and VRSQ equally revealed robust associations between them. Although the vestibular scores indicated weaker correlations compared to the other scores, the correlations were moderate (see Table 4). These outcomes further support the convergent and construct validity of both versions of the CSQ-VR.

**Table 4.** Correlations between the scores of CSQ-VR (both versions) and the VRSQ.

| Correlation Pair | | Pearson's r | p-value |
|---|---|---|---|
| CSQ-VR – Total Score | VRSQ – Total Score | 0.77 | < .001*** |
| CSQ-VR – Oculomotor | VRSQ – Oculomotor | 0.75 | < .001*** |
| CSQ-VR – Vestibular | VRSQ – Disorientation | 0.55 | < .001*** |
| CSQ-VR (VR) – Total Score | VRSQ – Total Score | 0.65 | < .001*** |
| CSQ-VR (VR) – Oculomotor | VRSQ – Oculomotor | 0.62 | < .001*** |
| CSQ-VR (VR) – Vestibular | VRSQ – Disorientation | 0.52 | < .001*** |

*(VR) = VR version*

### 3.2. Detection of Temporary Decline due to Cybersickness

As mentioned above, the temporary decline was defined by two standard deviations from the mean of the baseline assessment. This definition is in line with the guidelines of the American Academy of Clinical Neuropsychology for determining whether performance is abnormal [44]. Eleven observations were detected which met the criterion for temporary cognitive/motor decline. Six of these were pertinent to reaction speed (i.e., longer reaction times) and five of them were applicable to motor speed (i.e., slower). Thus, all temporary declines were found for the DLRT task. A trend was also observed where, when motor speed was substantially slower (i.e., a decline), then reaction speed also substantially declined. Finally, these declines were all found for three participants. Thus, only three participants experienced a temporary decline in their psychomotor skills. These results indicate that the susceptibility to experiencing a temporary decline due to cybersickness may be attributed to individual differences.

The ROC-AUC analyses provide cut-off scores for each questionnaire, where the optimal sensitivity (i.e., the detection of true positives) and specificity (i.e., the exclusion of true negatives) are achieved. Following the recommendations for ROC-AUC analyses and psychometrics [42], [43] to determine the suitability of a questionnaire to detect temporary decline, two criteria were set 1) AUC > 70% and 2) Metric Score > 1.5, which both had to be met. The ROC-AUC analysis for declines in reaction time (i.e., slower reaction times) showed that only the total scores for both versions of the CSQ-VR met the criteria (see

Table 5). Similarly, the ROC-AUC analyses for motor speed decline indicated that only the total scores for both versions of the CSQ-VR met the criteria. Furthermore, the two versions of the CSQ-VR showed the best sensitivity and specificity in detecting a temporary decline in reaction time and motor speed, while the VRSQ and SSQ showed significantly smaller psychometric properties (see Table 5, Table 6, and Figure 6). These results postulate that the total scores for both versions of the CSQ-VR have superior psychometric properties to the total scores of the SSQ and VRSQ. Also, only the CSQ-VR total scores are suitable for detecting a temporary decline in reaction speed and/or motor speed.

**Table 5.** Psychometric Properties of the CSQ-VR, SSQ, and VRSQ in detecting Reaction Speed Decline.

| Cybersickness Score | Cut-off | Sensitivity (%) | Specificity (%) | PPV (%) | NPV (%) | AUC (%) | Metric Score |
|---|---|---|---|---|---|---|---|
| CSQ-VR – Total Score | 10 | 100% | 75% | 15.15% | 100% | 87% | 1.75 |
| CSQ-VR (VR) – Total Score | 9 | 100% | 75% | 15.15% | 100% | 86.5% | 1.75 |
| SSQ – Total Score | 83.36 | 80% | 68.75% | 10.26% | 98.72% | 66.1% | 1.49 |
| VRSQ – Total Score | 20 | 100% | 53.57% | 8.77% | 100% | 66.6% | 1.54 |

*(VR) = VR version; Based on [42] and [43], the following thresholds were set and had to be met: AUC > 70% and Metric Score > 1.5.; PPV = Positive Predictive Value (i.e., the ratio of true positives); NPV = Negative Predictive Value (i.e., the ratio of true negatives).*

**Table 6.** Psychometric Properties of the CSQ-VR, SSQ, and VRSQ in detecting Motor Speed Decline.

| Cybersickness Score | Cut-off | Sensitivity (%) | Specificity (%) | PPV (%) | NPV (%) | AUC (%) | Metric Score |
|---|---|---|---|---|---|---|---|
| CSQ-VR – Total Score | 10 | 100% | 75.68% | 18.18% | 100% | 86.9% | 1.76 |
| CSQ-VR (VR) – Total Score | 9 | 100% | 75.68% | 18.18% | 100% | 88% | 1.76 |
| SSQ – Total Score | 83.36 | 83.33% | 69.37% | 12.82% | 98.72% | 68% | 1.53 |
| VRSQ – Total Score | 20 | 100% | 54.05% | 10.53% | 100% | 67.53% | 1.54 |

*(VR) = VR version; Based on [42] and [43], the following thresholds were set and had to be met: AUC > 70% and Metric Score > 1.5; PPV = Positive Predictive Value (i.e., the ratio of true positives); NPV = Negative Predictive Value (i.e., the ratio of true negatives).*

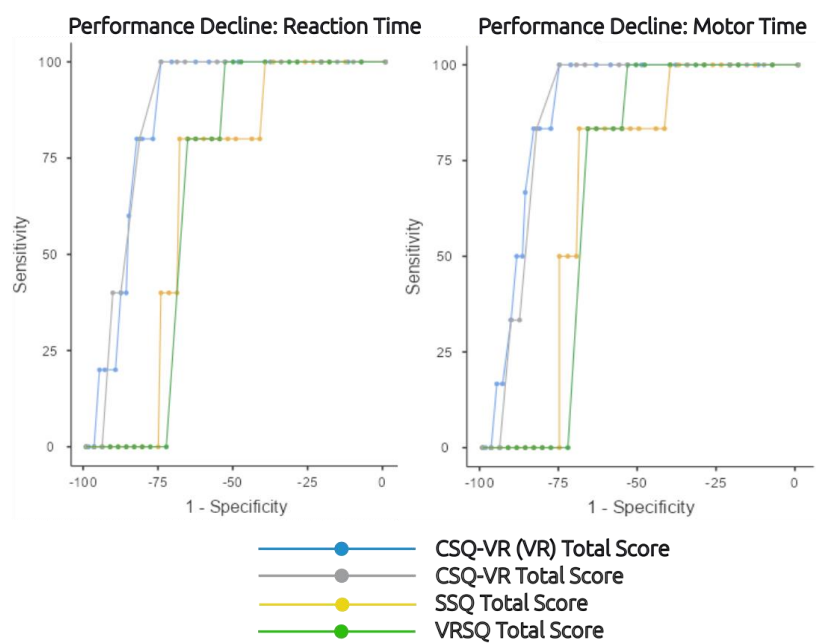

**Figure 6.** Sensitivity and Specificity of the CSQ-VR, SSQ, and VRSQ Total Scores in detecting Reaction Time (Left) and Motor Speed (Right) Decline. *Note: (VR) = VR version*

The psychometric properties of the sub-scores of each questionnaire were also examined. For detecting a temporary decline in reaction speed or motor speed, only the Vestibular/Disorientation scores of the questionnaires met the criteria of suitable psychometric properties (see Table 7 and Table 8). However, the Nausea score of the VR version of CSQ-VR also met the criteria for detecting both. The best sensitivity and specificity in detecting temporary decline of either reaction or motor speed was observed for the Vestibular score of the paper-and-pencil version of the CSQ-VR, closely followed by the same score for the VR version of the CSQ-VR (see Table 7, Table 8, and Figure 7). The sensitivity and specificity of the Disorientation scores of the SSQ and VRSQ were substantially lower compared to CSQ-VR. However, the sensitivity and specificity of the Disorientation score of the SSQ were significantly higher than the ones for the Disorientation score of the VRSQ.

**Table 7.** Psychometric Properties of the CSQ-VR, SSQ, and VRSQ Vestibular/Disorientation Scores in detecting Reaction Speed Decline.

| Cybersickness Score | Cut-off | Sensitivity (%) | Specificity (%) | PPV (%) | NPV (%) | AUC (%) | Metric Score |
|---|---|---|---|---|---|---|---|
| CSQ-VR – Nausea | 3 | 60% | 67.86% | 7.69% | 97.44% | 65.3% | 1.28 |
| *CSQ-VR – Vestibular* | *5* | *100%* | *77.68%* | *16.67%* | *100%* | *92.6%* | *1.78* |
| CSQ-VR – Oculomotor | 7 | 40% | 93.75% | 22.22% | 97.22% | 65.8% | 1.34 |
| *CSQ-VR (VR) – Nausea* | *3* | *100%* | *66.96%* | *11.09%* | *100%* | *83.6%* | *1.67* |
| *CSQ-VR (VR) – Vestibular* | *4* | *100%* | *70.54%* | *13.16%* | *100%* | *86.7%* | *1.71* |
| CSQ-VR (VR) – Oculomotor | 6 | 40% | 90.18% | 15.38% | 97.12% | 61.2% | 1.30 |
| SSQ – Nausea | 47.7 | 40% | 88.39% | 13.33% | 97.06% | 60.04% | 1.28 |
| *SSQ – Disorientation* | *11.22* | *100%* | *64.29%* | *11.11%* | *100%* | *70.1%* | *1.64* |
| SSQ – Oculomotor | 45.48 | 80% | 58.04% | 7.84% | 98.48% | 67.9% | 1.38 |
| *VRSQ – Disorientation* | *20* | *80%* | *74.01%* | *12.12%* | *98.81%* | *73.06%* | *1.54* |
| VRSQ – Oculomotor | 33.33 | 100% | 53.57% | 8.77% | 100% | 63.4% | 1.54 |

*(VR) = VR version; Based on [42] and [43], the following thresholds were set and had to be met: AUC > 70% and Metric Score > 1.5; PPV = Positive Predictive Value (i.e., the ratio of true positives); NPV = Negative Predictive Value (i.e., the ratio of true negatives).*

**Table 8.** Psychometric Properties of the CSQ-VR, SSQ, and VRSQ Vestibular/Disorientation Scores in detecting Motor Speed Decline.

| Cybersickness Score | Cut-off | Sensitivity (%) | Specificity (%) | PPV (%) | NPV (%) | AUC (%) | Metric Score |
|---|---|---|---|---|---|---|---|
| CSQ-VR – Nausea | 2 | 100% | 32.43% | 7.41% | 100% | 62.6% | 1.32 |
| *CSQ-VR – Vestibular* | *5* | *100%* | *78.38%* | *20%* | *100%* | *94.4%* | *1.78* |
| CSQ-VR – Oculomotor | 7 | 33.33% | 93.69% | 22.22% | 96.3% | 61% | 1.27 |
| *CSQ-VR (VR) – Nausea* | *3* | *100%* | *67.57%* | *14.29%* | *100%* | *85.1%* | *1.68* |
| *CSQ-VR (VR) – Vestibular* | *4* | *100%* | *71.17%* | *15.79%* | *100%* | *89.3%* | *1.71* |
| CSQ-VR (VR) – Oculomotor | 6 | 33.33% | 90.09% | 15.38% | 96.15% | 56.5% | 1.23 |
| SSQ – Nausea | 47.7 | 50% | 89.19% | 20% | 97.06% | 65.08% | 1.39 |
| *SSQ – Disorientation* | *11.22* | *100%* | *64.86%* | *13.33%* | *100%* | *70.3%* | *1.65* |
| SSQ – Oculomotor | 45.48 | 83.33% | 58.56% | 9.8% | 98.48% | 67.8% | 1.42 |
| *VRSQ – Disorientation* | *20* | *82.3%* | *74.77%* | *15.15%* | *98.81%* | *75%* | *1.58* |
| VRSQ – Oculomotor | 33.33 | 100% | 54.05% | 10.53% | 100% | 63.5% | 1.54 |

*(VR) = VR version; Based on [42] and [43], the following thresholds were set and had to be met: AUC > 70% and Metric Score > 1.5; PPV = Positive Predictive Value (i.e., the ratio of true positives); NPV = Negative Predictive Value (i.e., the ratio of true negatives).*

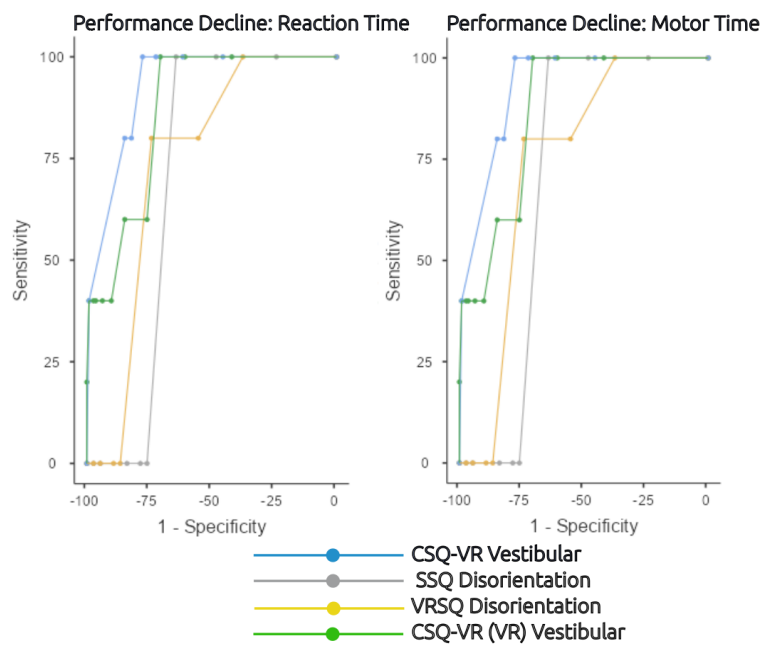

**Figure 7.** Sensitivity and Specificity of the CSQ-VR, SSQ, and VRSQ Vestibular/Disorientation Scores in detecting Reaction Time (Left) and Motor Speed (Right) Decline. *Note: (VR) = VR version*

*3.3. Mixed Model Regression Analysis*

A mixed model regression analysis was conducted to determine whether pupil size can be a biomarker/predictor of cybersickness. The analysis indicated that the model with pupil size as a predictor of the Total Score on the VR version of CSQ-VR was significant. Pupil size also revealed a relatively high beta (negative) coefficient, postulating that cybersickness intensity substantially increases as pupil size decreases (see Figure 8). Furthermore, the fixed effects of pupil size, alongside the random effects of the participants, appear to explain 50% of the variance for the intensity of cybersickness. These outcomes postulate that pupil size is a significant predictor of the intensity of cybersickness, and it therefore can be considered as a biomarker of cybersickness.

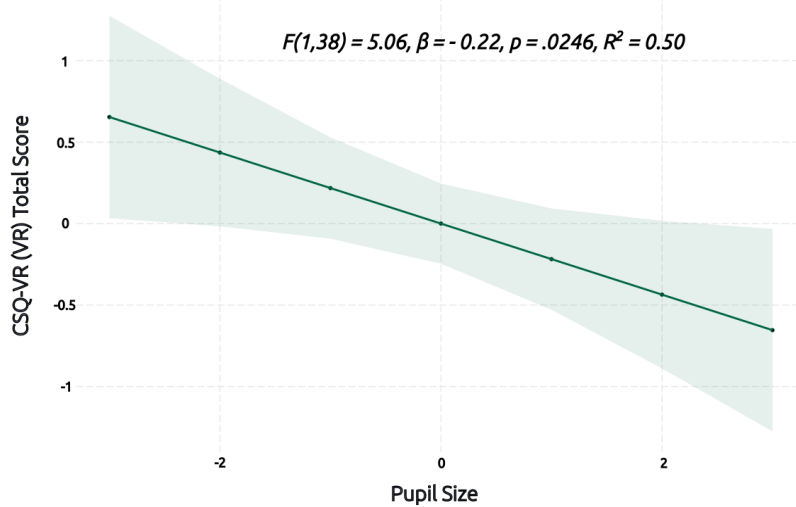

**Figure 8.** Mixed Regression Model of Pupil Size Predicting Cybersickness Intensity

## 4. Discussion

The CSQ-VR is an adapted and enhanced version of the VRISE section and sub-score of the VNRQ. Based on the recommendations by Ursachi et al. [41] for Cronbach's $\alpha$, the CSQ-VR displayed good to very good internal consistency. This finding is aligned with the high structural validity and internal consistency of the VRNQ, and its VRISE sub-score, which have previously been observed [29]. Also, the total scores and sub-scores of both versions of the CSQ-VR showed robust correlations with their respective sub-scores and total scores of the SSQ and VRSQ. This finding supports Somrak et al. [30] where the VRISE sub-score of the VRNQ was significantly correlated with the SSQ Total Score. Nevertheless, the current study meticulously examined the reliability and validity of the total scores and the sub-scores of both versions of the CSQ-VR. Beyond their convergent validity, the associations between the sub-scores of the CSQ-VR (both versions), Nausea, Vestibular, and Oculomotor, with the equivalent sub-scores of the SSQ, support the construct validity of both versions of the CSQ-VR for examining the whole range of cybersickness symptomatology. Therefore, both the paper-and-pencil and the VR versions of the CSQ-VR are highly reliable and valid tools for measuring the presence and intensity of cybersickness symptoms in VR.

*4.1. Comparison of CSQ-VR, SSQ, and VRSQ*

Several studies have reported that the SSQ does not have adequate psychometric properties for measuring cybersickness in VR [20], [51], [52]. The findings of this current study are aligned with the previous literature. The inadequacy and inappropriateness of the SSQ for measuring cybersickness in VR have also been confirmed. Specifically, the overall and sub-scores for the SSQ and the VRSQ displayed internal consistency which was substantially inferior to the respective internal consistency of the CSQ-VR total score and sub-scores. Moreover, the Nausea item of the SSQ revealed internal consistency that is below the parsimonious threshold of 0.7 for Cronbach's $\alpha$, which is required for a tool to be used in research and professional settings [53]. Likewise, the Oculomotor sub-score of the VRSQ was well below this threshold. Given that the VRSQ has only two sub-scores (i.e., Disorientation and Oculomotor), this means that half of the test was found to be unreliable. This finding agrees with the serious limitations reported in the VRSQ development and validation, which was conducted using a smartphone VR (i.e., Samsung Gear VR) and not a PC or standalone VR, a very simplistic task (i.e., target selection) and stimuli (i.e., small and large buttons) which were not efficient in inducing adequate levels of cybersickness, in a relatively small sample [21]. As a result, all the items pertinent to Nausea, which is the second most frequent symptom of cybersickness in VR [7], [22]–[24], [51], were dropped. Thus, it comes as no surprise that both the SSQ and VRSQ displayed problematic consistencies in certain sub-scores and overall inferior reliability for the total and sub-scores of the CSQ-VR.

Furthermore, the SSQ has received criticism for its highly complex structure and scoring [30], [51]. The CSQ-VR has previously been strongly preferred over the SSQ because of its easily calculated and interpretable scores [30]. The VRSQ, deriving from the SSQ, has predominantly maintained the SSQ structure and scoring system, albeit the VRSQ scoring system requires somewhat simpler calculations. Nevertheless, as was also seen in this study, both the SSQ and VRSQ suffer in terms of structure. Also, given that the design of the questions and available responses using a Likert scale is essential for collecting reliable and informative data [25]–[28], the CSQ-VR has an advantage over the SSQ and VRSQ. Both the SSQ and VRSQ use a 4-point Likert scale, while the relevant literature suggests that a 7-point Likert scale, especially when combining a number with textual information (e.g., *"6 – Very Intense Feeling"*), like in the CSQ-VR, are substantially more efficient in providing useful and representative self-reports [25]–[28]. The design of the general instructions (i.e., *"Please, from 1 to 7, circle the response that better corresponds to the presence and intensity of the symptom."*) and questions (e.g., *"Nausea A: Do you experience nausea (e.g., stomach pain, acid reflux, or tension to vomit)?"*) are also more explicit in the CSQ-VR than the equivalent design in the SSQ and VRSQ (i.e., general instruction: *"Circle how much each symptom below is affecting you now."*; question: *"Nausea"*). Finally, the SSQ has 16 questions measuring the whole range of symptoms, the VRSQ has 9 questions, but it measures only Vestibular and Oculomotor related symptoms, while the CSQ-VR is shorter, measuring the whole range of cybersickness with only 6 questions. Therefore, the CSQ-VR is a shorter questionnaire with an overall superior design to the SSQ and VRSQ, which was also reflected in the psychometric properties examined in this study.

The previous literature has shown that cybersickness, particularly when symptoms are strong, may affect the cognitive and/or motor skills of the user [12]–[14], especially their reaction speed [10], [17], [18]. It is thus assumed that a questionnaire designed to measure cybersickness would also be effective in detecting the relevant declines in performance. The total score for both versions of the CSQ-VR showed high sensitivity and specificity in detecting these temporary declines in performance due to cybersickness, while the psychometric properties of the total scores of the SSQ and VRSQ were substantially lower and inadequate. Furthermore, two sub-scores (Nausea and Vestibular) of the CSQ-VR were also highly sensitive and specific in the detection of temporary declines, while the equivalent scores of the SSQ and VRSQ (it does not include a Nausea score) were either significantly inferior or inadequate. The CSQ-VR thus is the only questionnaire, which is effective in detecting these temporary declines in performance modulated by cybersickness. Given that VR is implemented in education [1] and professional training [2], neuropsychological assessment [54] and therapy [4], where cognitive and motor skills should be reliable, it is essential that a tool should be able to provide information that these skills may have been compromised by cybersickness symptomatology. Finally, beyond these applications, VR is gradually becoming established as a research tool in scientific fields, such as human-computer interaction [55] and psychological sciences [3], where cybersickness may compromise the reliability of the scientific findings [12]. Thus, the detection of a participant whose performance has been compromised by cybersickness enables the exclusion of this participant or observation from the analyses and assures the data's reliability.

Nevertheless, as was also observed in this study, a participant's performance may not be affected throughout the experiment. In previous studies, changes in the intensity of cybersickness during exposure can occur in terms of increasing due to aggravation [22] or decreasing due to a cultivated tolerance [56]. Therefore, the continuous or repetitive assessment of cybersickness is required while the participant/user is immersed. Instead of excluding all of a participant's observations, the CSQ-VR allows a researcher to drop only those particular observations where a participant's performance was affected by cybersickness. Considering that the VR version of the CSQ-VR has shown comparable (and sometimes superior) psychometric properties to its paper-and-pencil version, it can detect specific compromised observations/performance and suggest its exclusion from the analyses. Nevertheless, beyond self-reports, there are other neuro- and bio-markers that have been used for detecting and measuring cybersickness [8]. Specifically, researchers have efficiently implemented electroencephalography [57], [58] and eye-tracking [59], [60] to detect and appraise the occurrence and intensity of cybersickness. The VR version of CSQ-VR benefits also from eye-tracking metrics. In this study, pupil size was found to be a significant predictor of cybersickness. A decrease in pupil size indicated higher intensity cybersickness, and vice versa, a pattern that has been previously observed between pupil size and negative emotions [31]. Previously, pupil size has been included in a deep fusion model for predicting cybersickness [60], however, its relationship and predictive ability and contribution to this model were not evaluated, preventing a conclusion of whether pupil size is a biomarker of cybersickness. This study provides evidence postulating that pupil size is indeed a biomarker of cybersickness, as well as its intensity. The VR version of CSQ-VR thus has an additional advantage of incorporating pupillometry.

*4.2. Limitations and Future Studies*

The current study also has limitations that should be considered. The sample consisted of young adults, which prevented the examination of cybersickness in a more age-diverse population. Future studies should attempt to examine cybersickness in a sample with a greater age spectrum to enable the study of age differences in tolerance and/or susceptibility towards cybersickness. Also, this study implemented a parsimonious inclusion criterion based on the MSSQ scores (i.e., excluding individuals who scored higher than the 75th percentile who could experience substantially more frequent and stronger cybersickness symptomatology). Given that the intensity and prevalence of cybersickness substantially differ across individuals, future studies should explore the effects of cybersickness on cognitive and motor skills in a sample that may experience stronger symptoms. Finally, the assessment included only working memory and psychomotor tests. Future studies should strive to examine more complex cognitive functions (e.g., episodic memory or decision-making) and motor skills (e.g., tasks that require fine motor skills and accuracy).

### 5. Conclusions

The CSQ-VR is a short, valid and reliable tool of cybersickness, which has superior psychometric properties to the SSQ and VRSQ. Also, the paper-and-pencil and the VR versions of the CSQ-VR were highly sensitive and specific in detecting temporary performance declines that were modulated by cybersickness. The VR version of the CSQ-VR provides further advantages by facilitating an assessment of cybersickness in the virtual environment, while the participant/user is immersed. Finally, the VR version of the CSQ-VR benefits from pupillometry (i.e., measurement of pupil diameter), which was found to predict the presence and intensity of cybersickness. Pupillometry may thus be applied in VR as a biomarker of positive (e.g., amusement) and negative (e.g., frustration) emotions.

**Supplementary Materials:** The Cybersickness in Virtual Reality Questionnaire (CSQ-VR) can be downloaded at: http://dx.doi.org/10.13140/RG.2.2.36571.03362

**Author Contributions:** Conceptualization, P.K., F.A., and S.M.; methodology, P.K., F.A., and S.M.; software, P.K.; validation, P.K., F.A. and S.M.; formal analysis, P.K.; investigation, J.L. and R.A.; resources, P.K.; data curation, J.L. and R.A.; writing—original draft preparation, P.K., J.L. and R.A.; writing—review and editing, P.K., F.A., and S.M.; visualization, P.K.; supervision, P.K. and S.M; project administration, P.K. and S.M.; funding acquisition, J.L. and R.A. All authors have read and agreed to the published version of the manuscript.

**Funding:** This research received no external funding

**Informed Consent Statement:** Informed consent was obtained from all subjects involved in the study.

**Data Availability Statement:** The data presented in this study are available on request from the corresponding author. The data are not publicly available due to the ethical approval requirements.

**Acknowledgments:** This work was funded by the School of Philosophy, Psychology and Language Sciences of the University of Edinburgh. The authors would like to thank the uCreate Studio of the University of Edinburgh for providing them with the VR equipment and tech support.

**Conflicts of Interest:** The authors declare no conflict of interest